\documentstyle[aps,prb,twocolumn]{revtex}
\begin{document}
\title{Magnetic imaging of the paramagnetic
Meissner effect in the granular high-T$_c$ superconductor
Bi$_2$Sr$_2$CaCu$_2$O$_x$}
\author{J.R. Kirtley$^a$, A.C. Mota$^b$, M. Sigrist$^c$, and
  T.M. Rice$^d$\\
$^a$ IBM Research, Yorktown Heights, NY 10598, USA \\
$^b$ Festk\"orperphysik, ETH-H\"{o}nggerberg, 8093 Z\"{u}rich,
Switzerland \\
$^c$ Yukawa Institute for Theoretical Physics, Kyoto University,
Kyoto 606-01, Japan \\
$^d$ Theoretische Physik, ETH-H\"{o}nggerberg, 8093 Z\"{u}rich,
Switzerland \\
}
\maketitle

\begin{abstract}
We have imaged the spatial distribution of magnetic flux on a
granular sample of the high-temperature superconductor
 Bi$_2$Sr$_2$CaCu$_2$O$_x$ using a scanning SQUID microscope.
Our results establish the presence of spontaneous orbital magnetic
moments which were suggested to be the origin of the paramagnetic
response of these materials. The signature of the orbital magnetic
moments is a rather broad distribution of local magnetic fields at the
surface of the sample. A simple model for the distribution is
presented.
\end{abstract}

\vskip 0.5 cm

%\narrowtext

Conventional superconductors generally tend to expel a small external
magnetic field upon cooling into the superconducting state. This
Meissner effect leads to complete, or
(due to remnant trapped flux, e.g. in a ceramic sample
composed of grains and voids) partial,
diamagnetism.
Therefore it came as a surprise when a paramagnetic signal
was observed in ceramic Bi$_2$Sr$_2$CaCu$_2$O$_x$ (Bi-2212)
\cite{svendlindh,lee,lan,braunisch}. The origin of the paramagnetism has been
a controversial subject. Braunisch {\it et al.} \cite{braunisch} and
Kusmartsev\cite{kusmartsev} proposed that some form of spontaneous
orbital currents were responsible, giving rise to magnetic moments
which could be aligned by the external field.
 This proposal for spontaneous orbital currents (Wohlleben Effect)
 in turn led two of us \cite{sigrist,sigrist2} to propose that an intrinsic $
d_{x^2-y^2} $-wave symmetry of the superconducting state would
naturally lead to frustrated Josephson junction circuits in a ceramic
sample where randomly oriented grains contact each
other\cite{geshkenbein}. Although
spontaneous orbital currents have now been unequivocally demonstrated
for high-temperature superconductors in controlled geometries, and the
evidence for $ d_{x^2-y^2} $-wave symmetry is now overwhelming
\cite{wollman,brawner,triprl,wollman2,miller,mathai,kouznetsov}, the
controversy has continued. In part this is due to the observation of
paramagnetic signals under quite different conditions, e.g. in bulk Nb
samples\cite{minhaj,kostic,rice,kostic2}.  In this letter we report
the first direct imaging of the local magnetic flux distribution in
the ceramics by a scanning SQUID microscope (SSM) and demonstrate
that a polarization of the distribution of spontaneous fluxes is
indeed responsible for the paramagnetic signal.

Before proceeding to a discussion of our experiment we would like to
remark that these two forms of paramagnetism in ceramic Bi-2212 and in
bulk Nb sample can be clearly distinguished in several other ways.
For example, the cooling rate affects the
magnetic response differently in the two cases.
Recent experimental data show
significant differences between
Nb and granular Bi$_2$Sr$_2$CaCu$_2$O$_x$ (Bi2212) samples.
While slow cooling enhances the paramagnetic
signal for the granular sample, it is diminished in the Nb sample. This
clearly indicates that the equilibrium state of both samples in a
small magnetic field is quite different\cite{gross}.
For the Nb-disks Koshelev and Larkin gave an explanation based on the
idea that during the cooling process the surface region nucleates
superconductivity before the bulk, so that magnetic flux in the sample
is compressed and creates an enhanced
magnetization\cite{koshelev}. This compressed
flux mechanism leads to a metastable state which depends on the
cooling procedure whereas the polarization of the spontaneous orbital
moments is an equilibrium process.
Further, noise measurements of the magnetization of Bi2212 give signals
which are
compatible with the presence of spontaneous orbital moments\cite{magnuson}.

We used a sample whose preparation, characterization and measurement
of the magnetization were reported previously
\cite{braunisch}. The magnetic images were made with a high-resolution
SSM\cite{ssmapl}.
This instrument uses a Nb-Al$_2$O$_3$-Nb low T$_c$
SQUID fabricated on a silicon substrate.
The substrate is polished to
a sharp tip spaced a few tens of microns from a well-shielded superconducting
pickup loop, which is an integral part of the SQUID. The SQUID substrate is
mounted on a flexible cantilever, oriented at a shallow angle, typically
20 degrees relative to the sample surface, and the
sample is scanned relative to the SQUID, with the tip of the substrate
in direct contact with the sample. For these measurements
the sample was polished to a mirror finish, and both
sample and SQUID were coated with a thin layer to protect
the SQUID substrate from abrasion.
We estimate, from fits of Abrikosov vortices imaged using
similar tip geometry and
SQUID and sample coating techniques, that the spacing between the
pickup loop and the surface of the superconducting sample is about 5$\mu$m.
The SQUID signal is proportional to
the magnetic flux through the loop area. The present images
were taken with a square pickup loop.
In this geometry a single
bulk Abrikosov vortex couples about 0.5$\Phi_0$  ($\Phi_0 = h/2e$)
through the $ 8.2\mu m \times 8.2 \mu m $ area of the
pickup loop located directly above it.

%\begin{figure}[p]
%\parbox{15cm}{\epsfig{file=f:mota1.ps,height=13.5cm,width=8.0cm} }
%\psfig{figure=f:mota1.ps,height=13.0cm}
%\caption{Scanning SQUID microscope images of a granular Bi2212
%sample, cooled and imaged in various fields. Each image has 512x512
%pixels,
%with 6 microns per pixel. The individual images are
%labeled by the cooling field, and by the maximal range of variation of the
%flux (in units of $ \Phi_0$) in each case. The outlined areas are regions of the
%images analyzed further in the histograms of Fig.2.}
%\end{figure}

The Fig. 1(a-f) shows a series of SSM images of
the Bi2212 sample which was cooled through the superconducting transition
temperature ($\approx$ 84K) at different values of an
externally applied magnetic field. The images were taken with the
field still applied and the sample and SQUID
immersed in liquid helium at 4.2K.
Each image is of a square area 3mm on a side.
The outlined square
shows the 480 $\mu$m $\times$ 480$\mu$m area
of the images used to generate
the histograms of Fig.2. We analyze only this area, because
our SQUID
has two magnetic field sensitive regions:
the pickup loop, and the hole in the superconducting ground plane
for the flux modulation coil\cite{ssmapl,footnote}.
The images were taken with the SQUID
oriented vertically, with the pickup loop towards the top of the images.
Therefore, when the pickup loop covers the region outlined, the modulation
hole senses areas well off the sample (below the bottom of the image),
and simply contributes a constant background signal to the image.

%\begin{figure}[p]
%\psfig{figure=f:mota2.ps,height=6cm}
%%\parbox{15cm}{\epsfig{file=f:mota2.ps,height=6cm,width=8.0cm} }
%\caption{Histograms of the distribution of occurrence of each SQUID flux value
%in the outlined areas of Fig.1, normalized so that the integral yields
%unity. Each
%panel is labeled by the cooling field in Gauss.
%The horizontal axes
%correspond to the flux through the SQUID
%at a particular spot on the sample, relative to the background with the
%pickup loop far from the sample, i.e. $ \Phi - \Phi_{ext} $.}
%\end{figure}

In Fig.1 we show the spatial distribution of the magnetic flux in and
around the sample. The grey contrast scale is chosen so that white
corresponds to the largest and black to the smallest (often
negative) flux value. In all cases the flux is plotted relative to
the flux introduced by the
external field, which sets the grey level away from the sample. One
overall feature observable by eye is the
difference between the paramagnetic magnetization (sample is brighter
than the background) at weak fields,
and a diamagnetic signal (sample is darker than
the background), in the pickup loop at strong fields.
For weak external fields the
inhomogeneity of the
magnetic flux is clearly visible and gives rise to a broad
distribution of the local fluxes.

%\begin{table}
%\begin{center}
%\caption{Average flux $ \Phi_{av} = \langle \Phi - \Phi_{ext} \rangle $
%  measured in the SSM and
%  standard deviation $ \delta \Phi = \langle ( \Phi -
%  \Phi_{av})^2 \rangle^{1/2} $ for different applied fields.}
%\begin{tabular}{|c|cc|}
%$ H_{ext} $ & $ \Phi_{av}/ \Phi_0 $ & $ \delta \Phi/\Phi_0 $ \\
%\hline
%10 mOe & 0.0157 & 0.0384 \\
%30 mOe & 0.0128 & 0.0386 \\
%100 mOe & 0.0299 & 0.0494 \\
%300 mOe & 0.0318 & 0.0675 \\
%1 Oe   & -0.0676 & 0.139 \\
%3 Oe   & -0.51   & 0.465 \\
%\end{tabular}
%\end{center}
%\end{table}

The flux distributions relative to the external flux measured
in the outlined square are shown as
histograms in Fig. 2.
The distribution is broad, as anticipated above, and the average
value indicates the overall response, which is paramagnetic for weak fields. In
Tab.1 we list the average flux $ \Phi_{av} - \Phi_{ext} $ and the standard
deviation $ \delta \Phi = \langle (\Phi - \Phi_{av})^2
\rangle^{1/2} $ ($ \Phi_{ext} $ denotes the flux of the external field
through the pickup loop).

%\begin{figure}[p]
%\psfig{figure=width.ps,height=8cm}
%%\parbox{15cm}{\epsfig{file=width.ps,height=8cm,width=8.0cm} }
%\caption{Standard deviation of the flux distribution $ \delta \Phi =
%\langle ( \Phi - \Phi_{av})^2 \rangle^{1/2} $ (circles,solid line) as
%a function of the
%external field. The average flux $ \Phi_{av} $ of the SSM measurement
%(diamonds,long dashed line) compares well with the appropriately scaled
%magnetization of
%the whole sample (x, dashed line).$^4$}
%\end{figure}

We first compare our SSM data with the total magnetization measurement
of the whole sample\cite{braunisch}.
Both show the same qualitative dependence of the
low-temperature magnetization on the applied field (see Fig.3). In
particular, the
sign of the magnetization changes for both measurements at the same
field value $ B_{ext} \approx 0.6 G $. This confirms that the SSM data,
scanning only a part of the sample, is typical of the
magnetization of the whole sample.
Let us now turn to the width of the flux distribution, i.e. the
standard deviation from the average flux. The field dependence of $
\delta \Phi $
indicates the existence of spontaneous flux at low external fields. If
the flux
observed were entirely due to flux trapped and compressed between and inside
the grains, then we expect that both $ \Phi_{av} $ and $ \delta \Phi $
would tend to zero in the zero-field limit. This is, however, not
the case as we illustrate in Fig.3, where we observe that the
zero-field limit of $ \delta \Phi $ is finite. This can be
readily interpreted if we assume that the flux distribution at low
fields is mostly due to spontaneous orbital currents for low external
fields, which can flow in either direction. Thus we expect to see an
inhomogeneous field pattern even at zero external field.
The broadening of the flux distribution with increasing field
can be understood as due to flux trapping and Meissner effect of
the grains. Generally more magnetic flux concentrates in the voids and
essentially little flux is trapped inside the grains. This leads to a enhancement of
the contrast in the flux values for large external fields and,
consequently, to a broader distribution.

For the low-field regime the flux distribution can be easily simulated
using a model for the boundary between many grains. Such a grain
boundary can be considered as a long Josephson junction and may be
described by a Sine-Gordon equation\cite{owen},

\begin{equation}
\frac{\partial^{2}\varphi}{\partial x^{2}} =
\frac{1}{\lambda_{J}^{2}}\sin{( \varphi(x)+\theta(x) )},
\end{equation}
where $ \varphi $ is the Josephson phase difference on the grain
boundary and $ \lambda_J $ is the Josephson penetration depth. The
presence of 0- and $ \pi
$-junctions enters through $ \theta(x) $, which assumes the values 0 or $
\pi $ as a function of the position $ x $ along the grain boundary.
This model is simulated on a system of length $ L $ using
$ N $ mesh points to determine $ \varphi(n) $ for fixed values of
$ \theta(n) $\cite{scalmod}. The local flux $ \Phi(n) $
between mesh point
$ n $  and $ n-1 $ is given by $\Phi(n) = \Phi_0 (\varphi(n)-\varphi(n-1))/2\pi$.
The external field is introduced via the boundary conditions at
the two ends of the junction $ (\varphi(N)-\varphi(N-1))/2\pi =
(\varphi(2)-\varphi(1))/2\pi = B_{ext} L/N\Phi_0 $.
Using a relaxation method\cite{scalmod}, we calculate $ \varphi $ while
gradually lowering the temperature by introducing decreasing values of $
\lambda_J $. In Fig.4 we show the flux distribution obtained for the
case of $ L = 100 $, $ N =1000 $, starting with $ \lambda_J = 40 $, which
is decreased by successive division by 2 to a final value of
0.156. The low external field is $ B_{ext} = 0.1 \Phi_0/Ld $ ($ d $ is
the magnetic width of the junction). We get a broad
distribution with a shape that is qualitatively similar to the
experiment. The phase $ \varphi $
has essentially a random walk like dependence on $ x $ so that the
histogram has an approximately Gaussian distribution.
Within this simple model we can describe the generation of spontaneous
flux and the interaction effects between flux lines. However, the
broadening of the flux distribution with increasing field
is not properly reproduced because
the contrast between trapped flux and the screening grains is not taken into
account.

%\begin{figure}[p]
%\parbox{15cm}{\epsfig{file=f:mota3.ps,height=8.0cm,width=8.0cm} }
%\psfig{figure=f:mota3.ps,height=8cm}
%\caption{Histogram for the small field limit of the Sine-Gordon model
%described in the text.}
%\end{figure}

In summary we would like to emphasize that the low-field data obtained
by an SSM on a granular Bi2212 sample are in very good qualitative
agreement with the previous magnetization measurements and, in addition,
provide direct evidence for the presence of spontaneous orbital
currents. On the other hand, independent evidence\cite{gross,magnuson}
indicates that the paramagnetic
signal in Nb samples has a different mechanism, likely
the one proposed by Koshelev and Larkin\cite{koshelev}.
It is not possible
from this SSM measurement alone to determine the origin of the orbital
currents. However, a series of previous experiments
suggest strongly
that grain boundaries with intrinsic $ \pi $-phase shifts are
appearing in these high-temperature superconductors and are very
likely responsible for the paramagnetic
response\cite{wollman,brawner,triprl,wollman2,miller,mathai,kouznetsov}.
In this sense,
the paramagnetic response (Wohlleben effect) in granular Bi2212 systems
is consistent in all aspects with an explanation based on
$ d_{x^2-y^2} $-wave pairing symmetry\cite{sigrist,sigrist2,higashitani}.

We are very grateful to the late D. Wohlleben for attracting our
attention to this effect and for providing us with a sample produced by
members of his group.
We would also like to thank M.B. Ketchen, M. Bhushan, and A.W. Ellis for
assistance in the design and fabrication of the microscope used in this
work, and
K.A. Moler and D.J. Scalapino for assistance in the development of the
computer program used for the simulation presented here. This work was
supported by the Swiss Nationalfonds.

%
% Place here the list of the references:
%

\begin{figure}
\caption{Scanning SQUID microscope images of a granular Bi2212
sample, cooled and imaged in various fields. Each image has 512x512
pixels,
with 6 microns per pixel. The individual images are
labeled by the cooling field, and by the maximal range of variation of the
flux (in units of $ \Phi_0$) in each case. The outlined areas are regions of the
images analyzed further in the histograms of Fig.2.}

\vspace{0.3in}
\caption{Histograms of the distribution of occurrence of each SQUID flux value
in the outlined areas of Fig.1, normalized so that the integral yields
unity. Each
panel is labeled by the cooling field in Gauss.
The horizontal axes
correspond to the flux through the SQUID
at a particular spot on the sample, relative to the background with the
pickup loop far from the sample, i.e. $ \Phi - \Phi_{ext} $.}

\vspace{0.3in}
\caption{Standard deviation of the flux distribution $ \delta \Phi =
\langle ( \Phi - \Phi_{av})^2 \rangle^{1/2} $ (circles,solid line) as
a function of the
external field. The average flux $ \Phi_{av} $ of the SSM measurement
(diamonds,long dashed line) compares well with the appropriately scaled
magnetization of
the whole sample (x, dashed line).$^4$}

\vspace{0.3in}
\caption{Histogram for the small field limit of the Sine-Gordon model
described in the text.}

\label{autonum}
\end{figure}

\begin{table}
\begin{center}
\caption{Average flux $ \Phi_{av} = \langle \Phi - \Phi_{ext} \rangle $
  measured in the SSM and
  standard deviation $ \delta \Phi = \langle ( \Phi -
  \Phi_{av})^2 \rangle^{1/2} $ for different applied fields.}
\begin{tabular}{|c|cc|}
$ H_{ext} $ & $ \Phi_{av}/ \Phi_0 $ & $ \delta \Phi/\Phi_0 $ \\
\hline
10 mOe & 0.0157 & 0.0384 \\
30 mOe & 0.0128 & 0.0386 \\
100 mOe & 0.0299 & 0.0494 \\
300 mOe & 0.0318 & 0.0675 \\
1 Oe   & -0.0676 & 0.139 \\
3 Oe   & -0.51   & 0.465 \\
\end{tabular}
\end{center}
\end{table}


\begin{thebibliography}{99}
\bibitem{footnote} Usually this extra pickup area
is not apparent, since it is 1.2 mm away from the loop,
well removed from the sample, where the magnetic fields are
slowly varying. However, in this experiment,
the scanned areas are relatively large (length 3 mm), so that
the background effect is visible in the images. The influence of
this additional pickup is
clearly seen, for example, as the dark shadow at the top of Fig.1f.
\end{thebibliography}
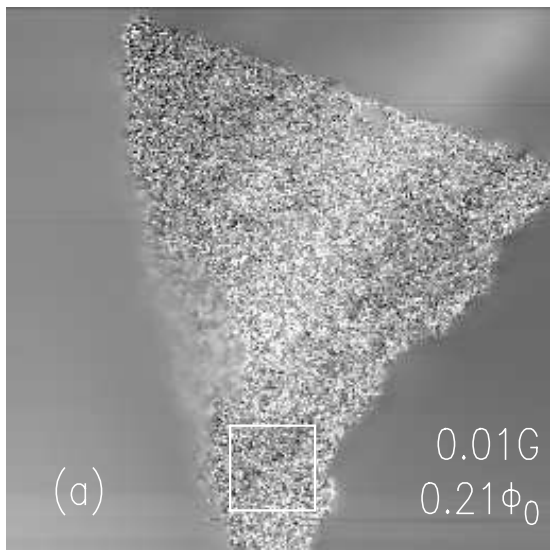
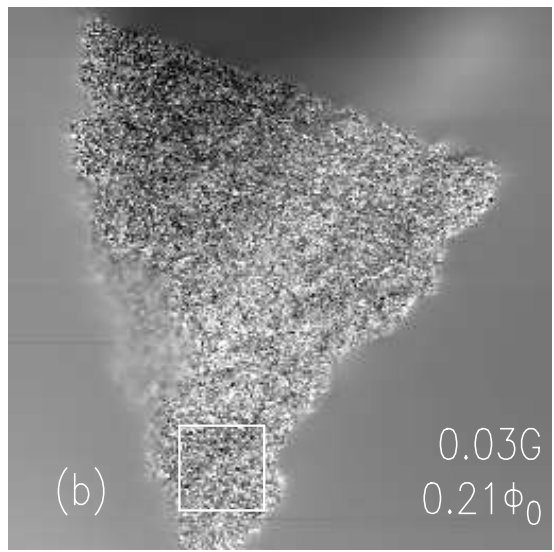
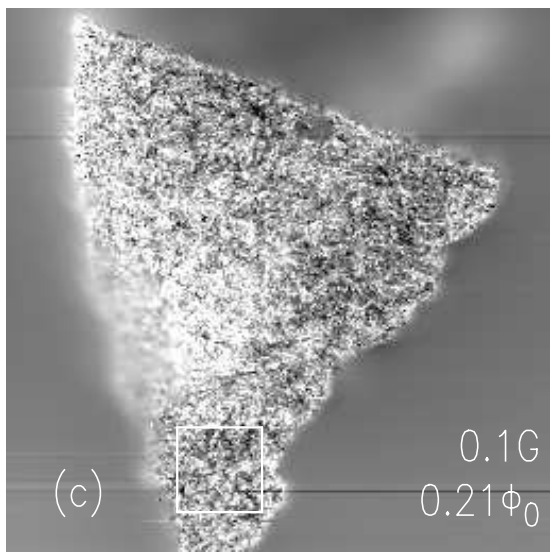
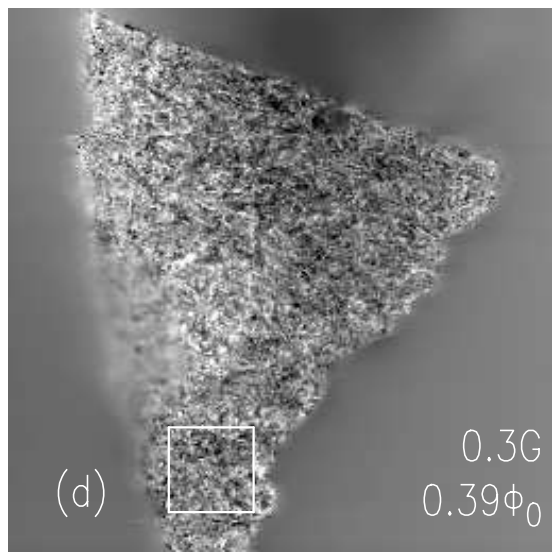
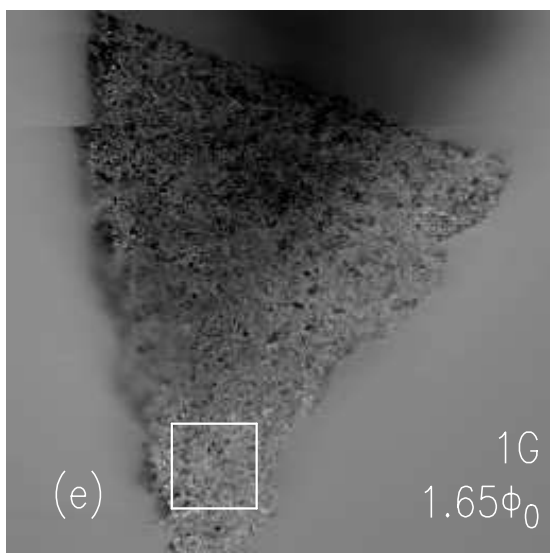
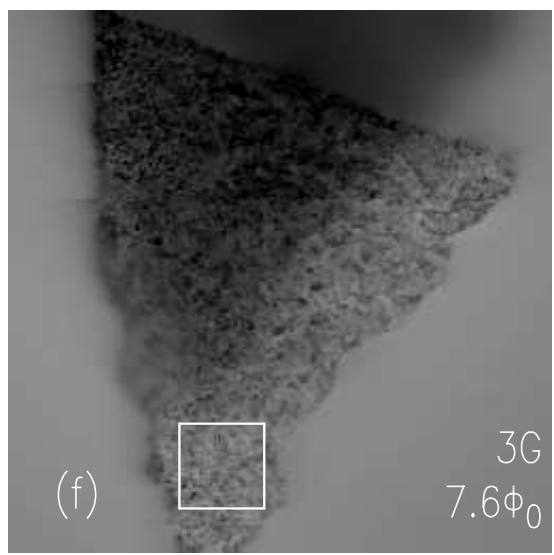
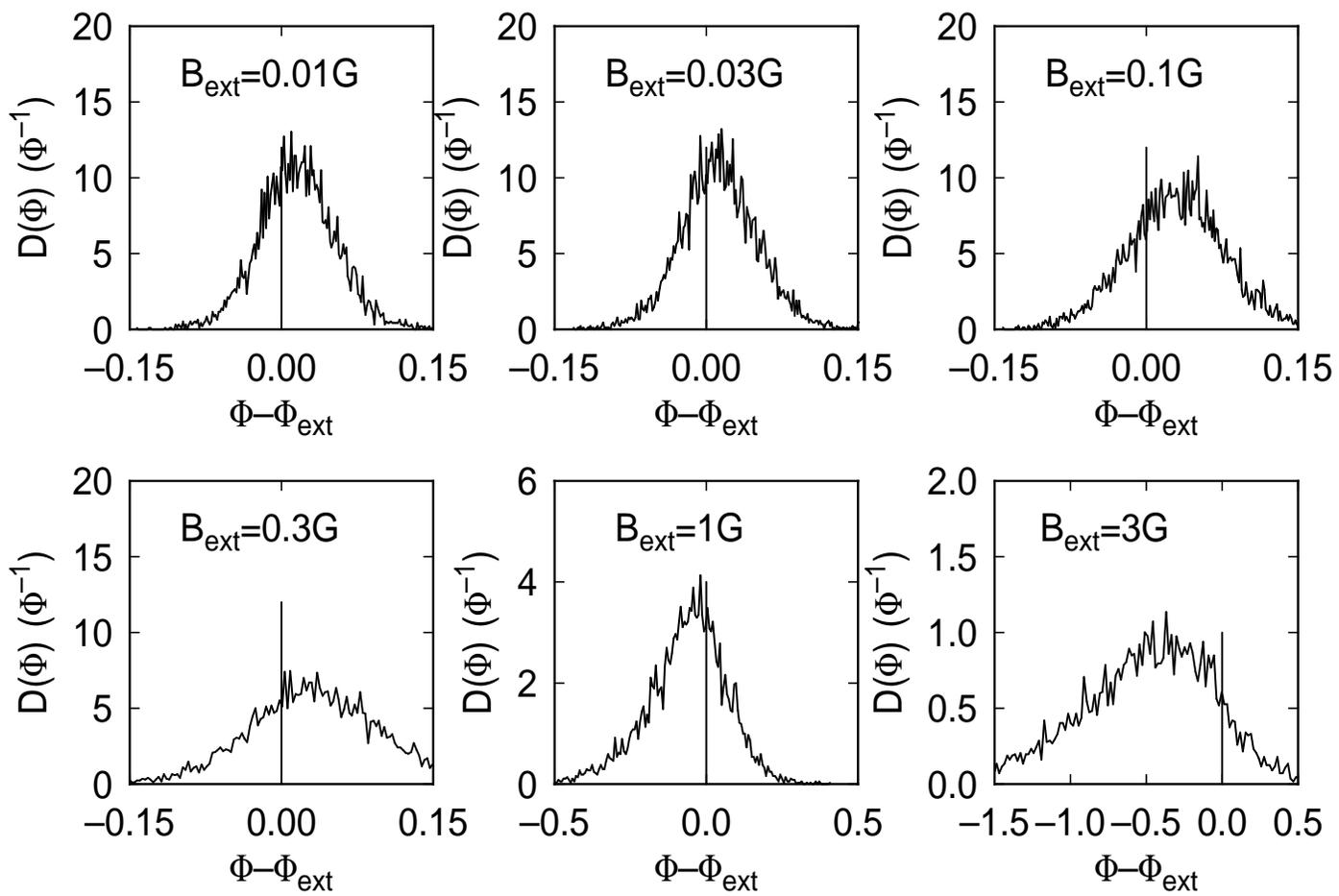
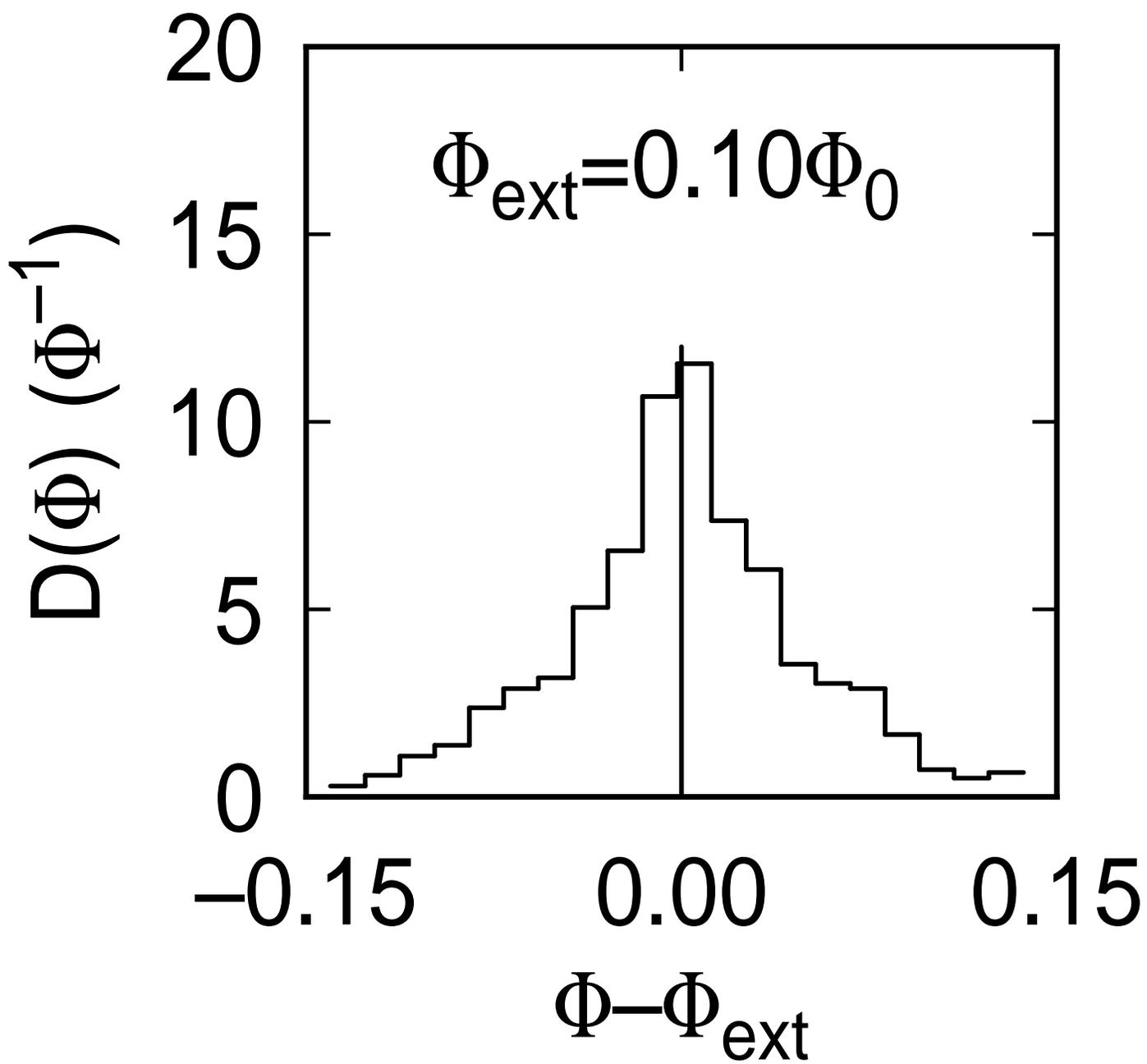
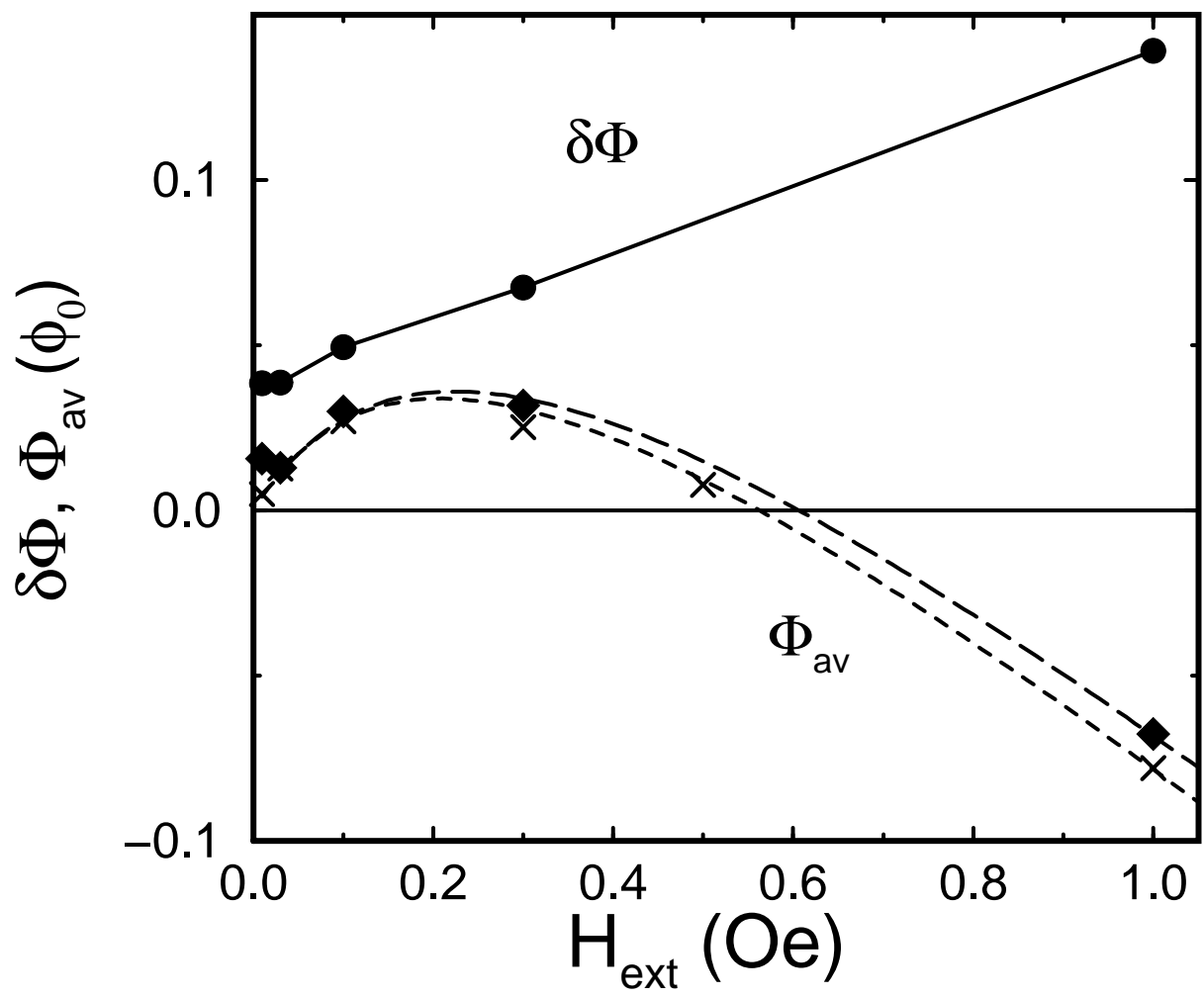
\end{document}